Inter-individual variability in sensory weighting of a plantar pressure-based, tongue-placed tactile biofeedback for controlling posture

Nicolas VUILLERME, Nicolas PINSAULT, Matthieu BOISGONTIER, Olivier CHENU, Jacques DEMONGEOT and Yohan PAYAN

Laboratoire TIMC-IMAG, UMR CNRS 5525, La Tronche, France

Number of text pages of the whole manuscript: 13

Number of figures: 2

Address for correspondence:

Nicolas VUILLERME

Laboratoire TIMC-IMAG, UMR CNRS 5525

Faculté de Médecine

38706 La Tronche cédex

France.

Tel: (33) (0) 4 76 63 74 86

Fax: (33) (0) 4 76 51 86 67

Email: nicolas.vuillerme@imag.fr

We would like this manuscript to be considered for fast communication (short communication)

**Abstract**

The purpose of the present experiment was to investigate whether the sensory weighting of a plantar pressure-based, tongue-placed tactile biofeedback for controlling posture could be subject to inter-individual variability. To achieve this goal, a large number of young healthy adults ($N = 45$) were asked to stand as immobile as possible with their eyes closed in two conditions of No-biofeedback and Biofeedback. Centre of foot pressure (CoP) displacements were recorded using a force platform. Overall, results showed reduced CoP displacements in the Biofeedback relative to the No-biofeedback condition, evidencing the ability of the central nervous system to efficiently integrate an artificial plantar-based, tongue-placed tactile biofeedback for controlling control posture during quiet standing. Results further showed a significant a positive correlation between the CoP displacements measured in the No-biofeedback condition and the decrease in the CoP displacements induced by the use of the biofeedback. In other words, the degree of postural stabilization appeared to depend on the each subject's balance control capabilities, the biofeedback yielding a greater stabilizing effect in subjects exhibiting the greatest CoP displacements when standing in the No-biofeedback condition. On the whole, by evidencing a significant inter-individual variability in sensory weighting of an additional tactile information related to foot sole pressure distribution for controlling posture, the present findings underscore the need and the necessity to address the issue of inter-individual variability in the fields of neuroscience and psychophysiology.

**Introduction**

We recently developed an original biofeedback system for improving balance whose underlying principle consists in supplying the user with supplementary sensory information related to foot sole pressure distribution through a tongue-placed tactile output device (Tongue Display Unit) (Vuillerme et al., 2006a). In a pioneering study, the effectiveness of this system in improving postural control during quiet standing has been evaluated on ten young healthy adults (Vuillerme et al., 2006a). By showing reduced centre of foot pressure (CoP) displacements when biofeedback was in use relative to when it was not, this experiment evidenced the ability of the central nervous system (CNS) to efficiently integrate an artificial plantar-based, tongue-placed tactile biofeedback for controlling posture during quiet standing. At this point, however, despite the overall stabilizing effect induced by the use of this biofeedback, and considering the existing literature reporting that individual subjects differ in the degree to which they weight visual (e.g. Chiari et al., 2000; Collins and De Luca, 1995; Crémieux and Mesure, 1994; Golomer et al., 1999; Isableu et al., 1997, 1998; Lacour et al. 1997; Vuillerme et al., 2001a), somatosensory (e.g. Fransson et al., 2000, 2003; Gurfinkel et al., 1995; Kluzik et al., 2005; Rogers et al., 2001; Vuillerme et al., 2001a,b; Wierzbicka et al., 1998), and vestibular (e.g. Fransson et al., 2003; Horak and Hlavacka, 2001; Wardman et al., 2003) information for controlling their balance, it is possible the sensory weighting of the tactile lingual cues for controlling posture to also be subject of inter-individual variability. Within this context, the present experiment was designed to evaluate this possibility, by testing a larger group of subjects.

**Materials and Methods**

Subjects

Forty-five young university student (age: $25.5 \pm 4.3$ years; body weight: $70.1 \pm 13.2$ kg; height: $177.6 \pm 10.7$ cm) participated in the experiment. They gave their informed consent to the experimental procedure as required by the Helsinki declaration (1964) and the local Ethics Committee, and were naive as to the purpose of the experiment. None of the subjects presented any history of motor problem, neurological disease or vestibular impairment.

Task and procedures

Subjects stood barefoot, in a natural position (feet abducted at 30°, heels separated by 3 cm), their arms hanging loosely by their sides with their eyes closed, and were asked to sway as little as possible. This postural task was executed under two experimental conditions of No-biofeedback and Biofeedback. The No-biofeedback condition served as a control condition. In the Biofeedback condition, subjects performed the postural task using a plantar pressure-based, tongue-placed tactile biofeedback system (Vuillerme et al., 2006a). A plantar pressure data acquisition system (FSA Inshoe Foot pressure mapping system, Vista Medical Ltd.), consisting of a pair of insoles instrumented with an array of $8 \times 16$ pressure sensors per insole (1cm² per sensor, range of measurement: 0-30 PSI), was used. The pressure sensors transduced the magnitude of pressure exerted on each left and right foot sole at each sensor location into the calculation of the positions of the resultant ground reaction force exerted on each left and right foot, referred to as the left and right foot centre of foot pressure, respectively ($CoP_{lf}$ and $CoP_{rf}$). The positions of the resultant CoP were then computed from the left and right foot CoP trajectories through the following relation (Winter et al., 1996):

$$CoP = CoP_{lf} \times R_{lf} / (R_{lf} + R_{rf}) + CoP_{rf} \times R_{rf} / (R_{rf} + R_{lf}),$$

where $R_{lf}$, $R_{rf}$, $CoP_{lf}$, $CoP_{rf}$ are the vertical reaction forces under the left and the right feet, the positions of the CoP of the left and the right feet, respectively.

CoP data were then fed back in real time to a recently developed tongue-placed tactile output device (Vuillerme et al., 2006a,b). This so-called Tongue Display Unit (TDU), initially introduced by Bach-y-Rita et collaborators (1998, 2003), comprises a 2D array (1.5 × 1.5 cm) of 36 electrotactile electrodes each with a 1.4 mm diameter, arranged in a 6 × 6 matrix. The matrix of electrodes, maintained in close and permanent contact with the front part of the tongue dorsum, was connected to an external electronic device triggering the electrical signals that stimulate the tactile receptors of the tongue via a flat cable passing out of the mouth. Note that the TDU was inserted in the oral cavity all over the duration of the experiment, ruling out the possibility the postural improvement observed in the Biofeedback relative to the No-biofeedback condition to be due to mechanical stabilization of the head in space. The underlying principle of our biofeedback system was to supply subjects with supplementary information about the position of the CoP relative to a predetermined adjustable "dead zone" (DZ) through the TDU. In the present experiment, antero-posterior and medio-lateral bounds of the DZ were set as the standard deviation of subject's CoP displacements recorded for 10 s preceding each experimental trial. A simple and intuitive coding scheme for the TDU, consisting in a "threshold-alarm" type of feedback was then used (Vuillerme et al., 2006a). (1) When the position of the CoP was determined to be within the DZ, no electrical stimulation was provided in any of the electrodes of the matrix. (2) When the position of the CoP was determined to be outside the DZ, electrical stimulation was provided in distinct zones of the matrix, depending on the position of the CoP relative to the DZ. Specifically, four different zones located in the front, rear, left and right of the matrix were defined; the activated zone of the matrix corresponded to the position of the CoP relative to the DZ. In the case that the CoP was located towards the front, rear, left or right of the DZ, distinct stimulations of the anterior, posterior, left and right zones of the matrix (i.e. stimulation of the front, rear, left or right portion of the tongue) were provided, respectively. Finally, in the present experiment, the frequency of the stimulation was maintained constant at 50 Hz across participants, ensuring a sensation of the continuous stimulation over the tongue surface. The intensity of the electrical stimulating current was adjusted for each subject, and for each of the front, rear, left and right portions of the tongue, given that the sensitivity to the electrotactile stimulation was reported to vary between individuals (Essick et al., 2003), but also as a function of location on the tongue in a preliminary experiment (Vuillerme et al., 2006b). Several practice runs were performed before starting the experiment to ensure that subjects had mastered the relationship between the position of the CoP relative to the DZ and lingual stimulations and were accustomed to the postural task.

A force platform (AMTI model OR6-5-1), which was not a component of the biofeedback system, was used to measure the displacements of the centre of foot pressure (CoP), as a gold-standard system for assessment of balance during quiet standing. Signals from the force platform were sampled at 100 Hz (12 bit A/D conversion) and filtered with a second-order Butterworth filter (10 Hz low-pass cut-off frequency).

Three 30s trials for each experimental condition were performed. The order of presentation of the two experimental conditions was randomized.

<u>Statistical analysis</u>

Since no statistically significant differences were observed between the three postural measurements recorded in the No-biofeedback condition and between the three postural measurements recorded in the Biofeedback condition, the mean values of the three trials performed in these two experimental conditions were used for statistical analyses. Two-tailed *t*-tests were then applied to the data from the No-biofeedback and Biofeedback conditions. In addition, Pearson's correlation coefficients (*r*) were calculated to assess whether the presumable decreased CoP displacements in the Biofeedback relative to the No-biofeedback condition were associated with subject's baseline postural control for standing in the No-biofeedback condition. To this aim, CoP data were normalized for each subject by calculating the percentage decrease in the surface area of the CoP measured in the Biofeedback condition compared to the No-biofeedback condition. These normalized decreases in surface area of the CoP were then plotted against the surface area of the CoP measured in the No-biofeedback condition. A significance level of $P<0.05$ was used for all tests.

**Results**

Analysis of the surface area of the CoP displacements showed smaller values in the Biofeedback than No-biofeedback condition ($t = 4.15$, $P<0.001$, Figure 1).

-----------------------------------------------
Please insert Figure 1 about here
-----------------------------------------------

Scatter plots of the surface area of the CoP displacements measured in the No-biofeedback condition vs. percentage decrease in the surface are of the CoP displacements measured in the Biofeedback relative to the No-biofeedback condition are illustrated in Figure 2. Analysis of surface area of the CoP showed a significant positive correlation between the surface area of the CoP measured in the No-biofeedback condition and the decrease in the surface area of the CoP displacements induced by the use of the biofeedback ($r = -0.51$, $P<0.001$, Figure 2).

-----------------------------------------------
Please insert Figure 2 about here
-----------------------------------------------

**Discussion**

The purpose of the present experiment was to identify whether additional sensory information related to foot sole pressure distribution to the user through a tongue-placed tactile output device could be weighted differently from one subject to another for controlling posture during quiet standing. To achieve this goal, a large number of young healthy adults ($N = 45$) were asked to stand as immobile as possible with their eyes closed in two conditions of No-biofeedback and Biofeedback. Centre of foot pressure displacements were recorded using a force platform.

On the one hand, results showed reduced CoP displacements in the Biofeedback than No-biofeedback condition (Figure 1). In general terms, this result confirms that an artificial tongue-placed tactile biofeedback can be efficiently integrated with other sensory cues by the postural control system to improve balance during quiet standing (Vuillerme et al., 2006a). At this point, one should keep in mind that the tongue was chosen as a substrate for electrotactile stimulation site according to its neurophysiologic characteristics. Indeed, because of its dense mechanoreceptive innervations (Trulsson and Essick, 1997) and large somatosensory cortical representation (Picard and Olivier, 1983), the tongue can convey higher-resolution information than the skin can (Sampaio et al., 2001; van Boven and Johnson, 1994). That is certainly one of the reason why the TDU already has proven its efficiency when used as the sensory output unit for tactile-vision (Bach-y-Rita et al., 2003; Sampaio et al., 2001) and tactile-proprioception (Robineau et al., 2006; Vuillerme et al., 2006b) sensory augmentation systems.

On the other hand, a significant inter-individual difference in the strength of the biofeedback on improving postural control during quiet standing was observed (Figure 2). At this point, we can exclude the possibility this observation differences to be due either to inter-individual differences in overall tactile sensitivity of the tongue (Essick et al., 2003) or to adaptation to the postural task. Indeed, (1) the intensity of the electrical stimulating current was adjusted to each subject (Vuillerme et al. 2006b) and (2) several practice runs were performed before starting the experiment to ensure that subjects had mastered the relationship between the position of the CoP relative to the DZ and lingual stimulations and were accustomed to the postural task and no detectable systematic changes were observed from the first to the last trial performed in each experimental condition. Rather, we propose the variability in the size of the stabilizing effect induced by the use of the biofeedback to stem to from the differences in how individual subjects have weighted the plantar-based, tongue-placed tactile biofeedback for controlling their posture during quiet standing. Note that such inter-individual variability is reminiscent of individual differences in postural responses observed consecutive to visual (e.g. Chiari et al., 2000; Collins and De Luca, 1995; Crémieux and Mesure, 1994; Golomer et al., 1999; Isableu et al., 1997, 1998; Lacour et al. 1997; Vuillerme et al., 2001a), somatosensory (e.g. Fransson et al., 2000, 2003; Gurfinkel et al., 1995; Kluzik et al., 2005; Rogers et al., 2001; Vuillerme et al., 2001a,b; Wierzbicka et al., 1998), and vestibular (e.g. Fransson et al., 2003; Horak and Hlavacka, 2001; Wardman et al., 2003) stimulations reported in previous experiments, hence lending support to the hypothesis that individual subjects differ in the degree to which they weight sensory information for controlling posture. Interestingly, results showed a significant positive correlation between the CoP displacements measured in the No-biofeedback condition and the decrease in the CoP displacements induced by the use of the biofeedback (Figure 2). In other words, the size of postural stabilization appeared to depend on the each subject's balance control capabilities, the biofeedback yielding a greater stabilizing effect in subjects exhibiting the greatest CoP displacements when standing in the No-biofeedback condition. With regard to the provision of additional tactile sensory information to the postural control system, our results corroborate those of a recent study in which the application of a tactile stimulus providing sway-related cues was reported to significantly reduce body sway, with the greatest percentage reduction in sway observed in subjects with the greatest sway while standing normally without any additional tactile stimulus (Rogers et al., 2001). A possible explanation for the greater reduction in sway by subjects that sway the most is that these subjects have a sensorimotor deficit so that the additional tactile stimuli are not as redundant (Rogers et al., 2001). Along these lines, in the context of the sensory reweighting mechanisms involved in postural control (e.g., Horak and Macpherson, 1996; Oie et al., 2002; Peterka and Loughlin, 2004), previous studies have shown that the sensory weights of each sensory input, in addition to vary from one individual to another, also depends on the sensory context (e.g. availability, accuracy, reliability, inconsistency between sensory signals). Indeed, to efficiently cope with changing environment/task/subject conditions, it is conceivable the CNS to adaptively and dynamically update the relative contributions of available sensory inputs to current conditions/constraints by (1) increasing the reliance on sensory modalities providing accurate and reliable information and (2) decreasing the reliance on sensory modalities providing inaccurate and unreliable information. At this point, it is possible that, for a given subject, the weight given to an augmented tactile input related to body sway for controlling posture during quiet standing to change according to the state/reliability of its somatosensory system, as observed by others (e.g. Dickstein et al., 2001; Rogers et al., 2001; Lackner et al., 2003; Vuillerme and Nougier, 2003). A study including experimental conditions of altered proprioceptive inputs from the leg muscles is currently being performed to address this issue.

**Conclusion**

In addition to the observation of an overall stabilizing effect of a plantar-pressure-based, tongue-placed tactile biofeedback (Vuillerme et al., 2006a), results of the present experiment showed this additional tactile sensory information for controlling posture during quiet standing to be weighted differently from one subject to another. On the whole, by evidencing a significant inter-individual variability in sensory weighting of an additional tactile biofeedback for controlling posture, the present findings underscore the need and the necessity to address the issue of inter-individual variability in the fields of neuroscience and psychophysiology. Indeed, inter-individual averaging, by definition, extracts only the collective effects that are coincident across subjects, hence failing to provide any information about individual variability in the functional organization of the CNS (e.g. Berthoz, 1997; Reuchlin et al., 1989).

**Acknowledgements**

The authors are indebted to Professor Paul Bach-y-Rita for introducing us to the TDU and for discussions about sensory substitution. The authors would like to thank subject volunteers. The company Vista Medical is acknowledged for supplying the FSA Inshoe Foot pressure mapping system.

**Figure captions**

**Figure 1.** Mean and standard deviation of surface area of the CoP displacements measured in the two No-biofeedback and Biofeedback conditions. These two experimental conditions are presented with different symbols: No-biofeedback (*white bars*) and Biofeedback (*black bars*). The significant $P$-value for comparisons between No-biofeedback and Biofeedback conditions also is reported (***: $P<0.001$).

**Figure 2.** Scatter plots of the surface area of the CoP displacements measured in the No-biofeedback condition *vs.* percentage decrease in the CoP displacements measured in the Biofeedback relative to the No-biofeedback condition. The significant positive correlation indicates that the biofeedback has a greater stabilizing effect in subjects exhibiting the greatest CoP displacements when standing in the No-biofeedback condition.

**Figure 1**

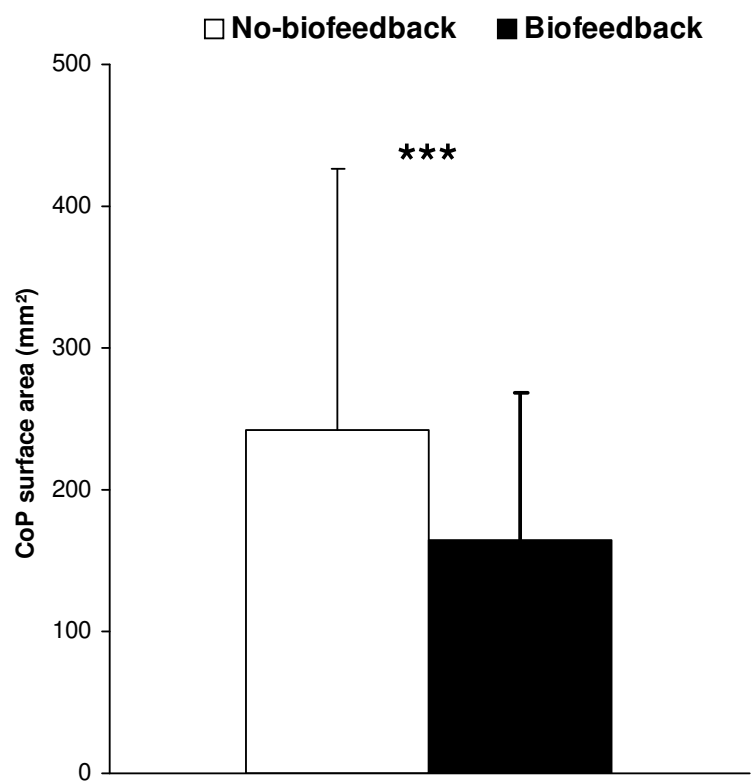

**Figure 2**

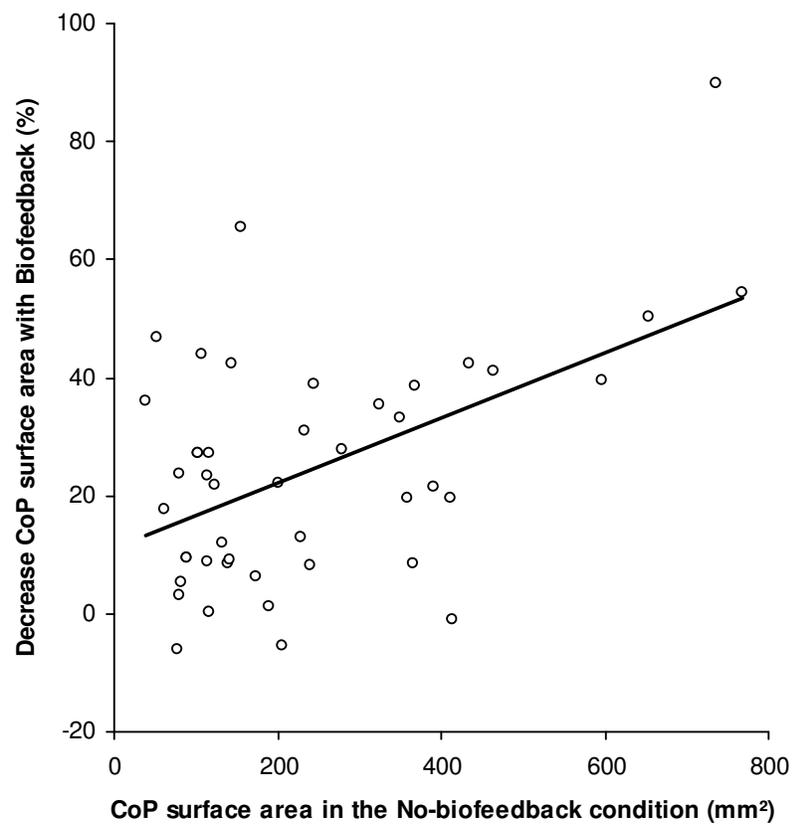